\documentclass[a4paper]{aa}
\usepackage{aabib, epsfig}

\def \deg{$^{\circ}$}

\begin{document}

\title{Towards a model of full-sky Galactic synchrotron intensity and
linear polarisation: a re-analysis of the Parkes data}

\author{G.~Giardino\inst{1},  
	A.~J.~Banday\inst{2},
  	K.~M.~G\'orski\inst{3,4}, 
	K.~Bennett\inst{1},
	J.~L.~Jonas\inst{5},
	J.~Tauber\inst{1}}

\institute{Research and Science Support Department of ESA, ESTEC,
  Postbus 299, NL-2200 AG Noordwijk, The Netherlands
\and
Max-Planck Institut f{\"u}r Astrophysik, Garching bei M\"{u}nchen,
                                             D-85741, Germany 
\and
ESO, Garching bei M\"{u}nchen, D-85748, Germany
\and
Warsaw University Observatory, Warsaw, Poland
\and
{Department of Physics \& Electronics, Rhodes University, PO Box 94,
  Grahamstown 6140, South Africa}
}

\offprints{G. Giardino} \mail{ggiardin@rssd.esa.int}

\date{Received date ; accepted date}

\titlerunning{Towards a model ...}
\authorrunning{G. Giardino et al.}

\abstract{We have analysed the angular power spectra of the Parkes
radio continuum and polarisation survey of the Southern galactic plane
at 2.4 GHz. We have found that in the multipole range $l=40$--$250$
the angular power spectrum of the polarised intensity is well
described by a power-law spectrum with fitted spectral index
$\alpha_{L} = 2.37 \pm 0.21$. In the same multipole range the angular
power spectra of the $E$ and $B$ components of the polarised signal
are significantly flatter, with fitted spectral indices respectively
of $\alpha_{E} = 1.57 \pm 0.12$ and $\alpha_{B} = 1.45\pm
0.12$. Temperature fluctuations in the $E$ and $B$ components are
mostly determined by variations in polarisation angle.  We have
combined these results with other data from available radio surveys in
order to produce a full-sky toy model of Galactic synchrotron
intensity and linear polarisation at high frequencies ($\nu \ga
10$~GHz). This can be used  to study the
feasibility of measuring the Cosmic Microwave Background polarisation
with forthcoming experiments and satellite missions.  } \maketitle

\section{Introduction}
\label{sec:intro}

Current cosmological models predict the level of polarisation of the
Cosmic Microwave Background (CMB) to be at 5--10\% of the temperature
anisotropies at the 20 arcmin angular scale, that is at $\mu$K level
(e.g. \cite{Bond87}). These temperature variations have to be detected
over a 3 K continuum with instrumentation whose typical noise
temperature is of the order of tens of Kelvin. The detection of the
CMB polarisation represents therefore a considerable challenge.  To
date, only upper limits on the CMB polarisation exist. The most recent
results from ground based observations place an upper limit of $10$
$\mu$K on any polarised signal from the sky at the 7\deg~ angular scale
in the frequency band 26--36 GHz (\cite{Keating01}) and an upper
limit of 15 $\mu$K at the 1\deg-0\deg.5 angular scale at $90$ GHz
(\cite{Hedman01}).

Current cosmological models also predict a shape for the angular power
spectra of the polarised component of the CMB, so the measure of these
spectra will provide additional information on the properties of the
primordial density fluctuations and on the thermal history of the
universe (e.g. \cite{Kosowsky96}).  The variation of polarised CMB
emission is predicted to peak at sub-degree angular scale.  The NASA
satellite mission MAP will have enough sensitivity to detect the CMB
polarisation, if this is present at the $\mu$K level (\cite{Kogut00}).
The future ESA mission Planck will have higher sensitivity and may
provide direct imaging of the CMB polarised signal
(\cite{Mandolesi98}).  The future SPORT experiment that is planned to
be placed on the International Space Station also has the sensitivity
to detect $\mu$K-level polarised signal from the CMB if this is
present at angular scales $\ga 7^{\circ}$ (\cite{Fabbri99}).

A fundamental question for any measurement of the polarisation of the
CMB is whether the expected signal can be distinguished from the
foreground polarised signal from our galaxy. The interstellar magnetic
field is illuminated by cosmic-ray electrons which spiral around
field lines and thereby emit synchrotron radiation. This radiation is
intrinsically highly linearly polarised, 70--75\% in a completely
regular field (e.g. \cite{rybicki}). Thermal emission from dust may
also be highly polarised, depending on the shape and alignment of the
dust particles (\cite{Wright87}; \cite{Prunet98}).  

As in the case of the unpolarised emission, knowledge of the
angular power spectra of the foreground polarised component is
essential when trying to recover the angular power spectra of the
CMB.  In this paper we use the Parkes 2.4 GHz polarimetric survey of
the southern Galactic plane in order to derive the angular power
spectra of the polarised component of the Galactic synchrotron
emission. The survey has relatively small sky coverage but has
10 arcmin angular resolution, which allows the statistical properties
of polarised synchrotron emission to be investigated at the angular
scales where the CMB polarised signal is expected to peak.


In order to assess the feasibility of CMB polarisation measurements,
study the effect of instrumental systematics and devise the data
analysis strategy it is necessary to perform simulations of the
satellite missions.  For these simulations an input sky with all the
known components of the emission is necessary. By exploiting existing
observations of the microwave sky, we have to strive
to construct a simulated sky which represents our best guess of some
of the properties of the real sky in the unexplored regions and
frequency channels. This will allow us to perform more realistic
simulations of the mission and to be in a better position to interpret
the real data when these become available.

We use the results of our analysis of the Parkes survey, combined with
other data on synchrotron emission coming from total intensity large
sky radio surveys, to produce a full-sky synthetic map of synchrotron
polarised emission. 
The paper is organised as follows. In Sect.~\ref{sec:duncun} the global
angular power spectra of the Parkes 2.4 GHz polarimetric survey are
derived.  The simulations performed to assess the reliability of the
spectra are presented in Sect.~\ref{sec:sim}. The results are
discussed in Sect.~\ref{sec:disco}.  In Sect.~\ref{sec:toys} the
derived  spectra are used as basis to construct a toy model of
the all-sky polarised synchrotron emission.

\section{The Parkes data at 2.4 GHz}
\label{sec:duncun}

The Parkes 2.4 GHz survey is a sensitive, polarimetric survey of the
southern Galactic plane carried out by \cite*{Duncan97}.  The
survey covers 127\deg~ of Galactic longitude (238\deg$\le l
\le$5\deg), with a latitude coverage out to at least $\pm$5\deg~ (up to
$b=$7\deg~ and $b=$8\deg~ over some longitudes). The angular resolution
of the images is 10.4 arcmin. The nominal rms noise of the survey
(for the total and polarised intensities) is 5.3 mK. The survey is publicly
available and the data come split into 6 rectangular
fields sampled at a pixel resolution of 4 arcmin. The linear
polarisation signal is given in terms of the two Stokes parameters $Q$
and $U$.

The Parkes survey has been recently analysed by
\cite*{Baccigalupi01} and \cite*{Tucci00}.  In both of these previous
studies, the data were split into twelve square patches, each of size
10\deg$\times$10\deg, which were then individually analysed. 
Our work differs from these in that the 
power spectra of the different polarisation components
are derived from a global analysis of the complete 
Parkes data.

In order to  perform this global statistical analysis of the
survey we have resampled the six fields of the original data into
HEALPix tessellations.  The iso-Latitude property of the HEALPix
tessellation of the sphere makes it ideal for fast computation of the
angular power spectrum of a field defined on the sphere
(\cite{Gorski98}). In the HEALPix tessellation the angular size of the
pixels is determined by the {\em nside} parameter.  We used a HEALPix
tessellation with {\em nside}=1024 which corresponds to pixels of
linear size 3.4 arcmin. When resampling the survey, the pixel value in
the HEALPix tessellation was assigned by weighted integration of the
values of the original pixels that overlapped with the new pixel,
where the weight of each pixel contribution is the fraction of its
area intersecting with the new pixel. This approach ensures that no
spurious pixel-to-pixel signal correlation is introduced by the
resampling.

\begin{figure*}[]. 
 \begin{center}
   \leavevmode \epsfig{file=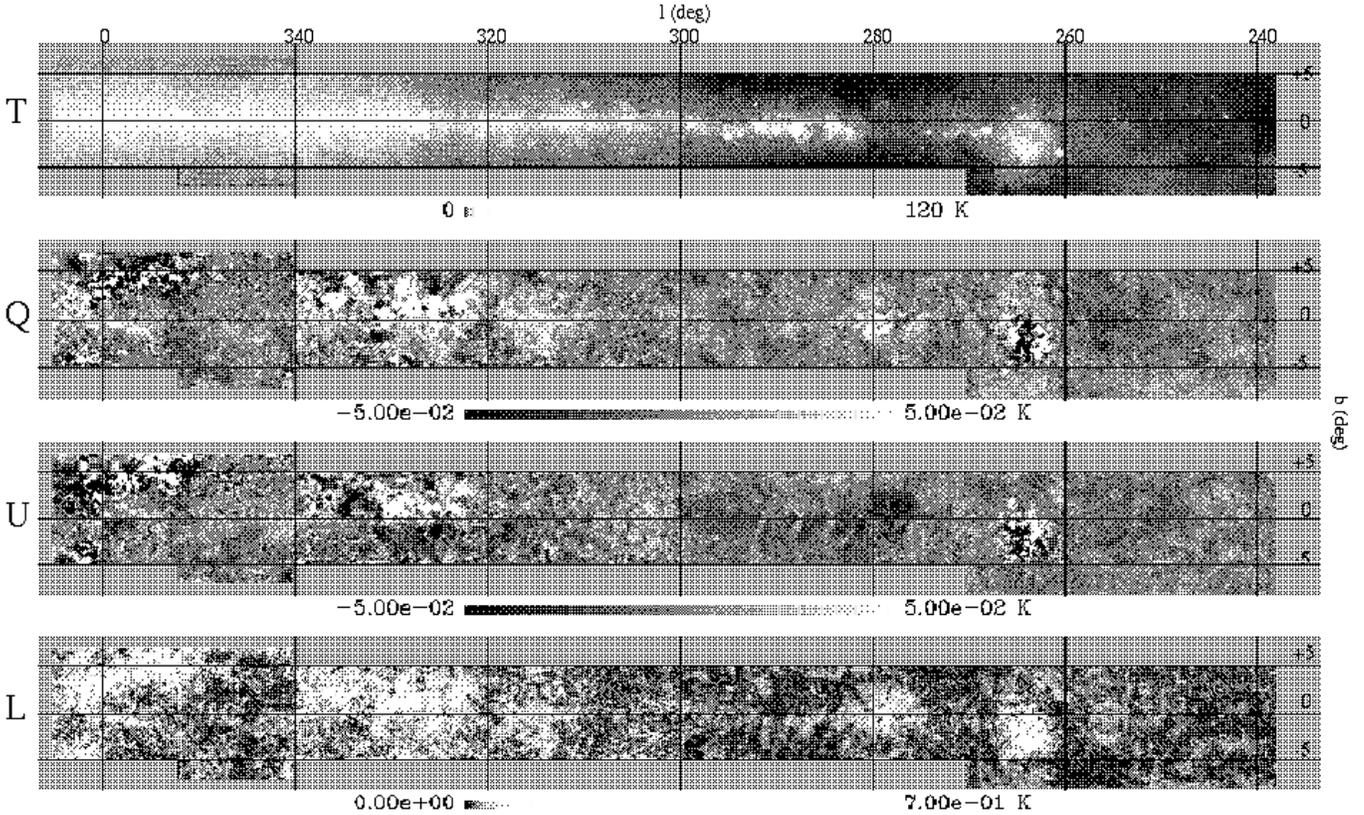,angle=0, width=\textwidth}

     \caption{The Parkes survey at 2400 MHz. The six fields of the
   original data have been resampled into one HEALPix tessellation
   with a pixel size of 3.4 arcmin. The first three images (from the
   top) give the three Stokes parameters of the signal: $T$, $Q$, $U$,
   respectively; the lowest image shows the derived polarised
   intensity $L = \sqrt{Q^2+U^2}$. The temperature scales for $T$ and
   $L$ have been histogram equalised.}
	
     \label{fig:parkes} 
 \end{center}
\end{figure*}

The three channels of the survey ($T$, $Q$ and $U$), resampled into
HEALPix are shown in Fig.~\ref{fig:parkes}. The  derived intensity
of linear polaristion, $L = \sqrt{Q^2+U^2}$, is also shown. The
maps are in Galactic coordinates.

The most striking difference between the total intensity channel ($T$) and the
polarisation channels ($Q, U$ and $L$) is that the total intensity
signal is dominated by the signal from the Galactic plane while in the
polarisation channels the Galactic plane is hardly visible.
Moreover many point sources which are clearly visible in the total intensity
channel do not exhibit a polarised counterpart. 

In the polarised intensity image the most apparent features are the
extended areas of polarisation, with sizes of the order of 5\deg~ or
so, which do not appear to be connected with discrete sources of total
power emission. Such regions lie around longitudes 265\deg, 280\deg,
320\deg~ to 330\deg~ and 355\deg~ to 5\deg. While one of these structures
can be identified with the Vela supernova remnant at $l=265$\deg, the
origin and nature of the other features is not yet explained
(\cite{Duncan97}).

Five regions that are bright in the $T$ maps have been blanked out to
prevent a significant fraction of instrumental polarisation from being
introduced. These are Sgr A and the bright H\,{\sc ii} complexes
G353.2+0.8, G284.14$-$0.3, G287.5$-$0.6 and G291.28$-$0.71.

\subsection{Derivation of the power spectra}

The three Stokes parameters $T$, $Q$ and $U$ fully describe any state
of linearly polarised light. However, whereas the temperature is a
scalar quantity (invariant under rotation of the plane of the sky),
$Q$ and $U$ are not. They depend on the direction of observation $\hat
n$ and on the Cartesian axes perpendicular to $\hat n$ used to define
them.  Recent theoretical developments have shown that it is more
useful to describe the polarisation field in terms of two quantities
that are invariant under rotation (\cite{Zaldarriaga97};
\cite{Kamionkowski97}). These two quantities are usually called $E$
and $B$ and they are obtained by a linear, non-local, transformation
of the $Q$ and $U$ map (hereafter we will refer to this transformation
as to the E-B transform). The $E$ and $B$ components differ in their
behavior under parity transformation: $B$ changes sign while $E$ does
not.  Four power spectra are needed to characterise the fluctuations
in a Gaussian theory: the autocorrelation of each of $T$, $E$ and $B$
and the cross correlation between $E$ and $T$.

Current cosmological models predict the CMB intensity and polarisation
variations to be a Gaussian field and therefore give predictions for
the four power spectra $T$, $E$, $B$ and $E\times T$. Galactic diffuse
emission is not a Gaussian field and so cannot be fully characterised
by the four power spectra, however, since we are interested in
comparing the statistical properties of this emission with predictions
for the CMB power spectra we have derived the four power spectra of
$T$, $E$, $B$ and $E\times T$ of the Parkes data.

The power spectra have been obtained by performing a spin-weighted
harmonic analysis of the field in the HEALPix grid with the ANAFAST
program (the software is part of the HEALPix package). The results of
the analysis for the whole set of data are summarised in the following
subsections. The angular power spectra of the polarised intensity was
derived by simple spherical harmonic decomposition, also using the
ANAFAST program.

From the analysis of large sky surveys, the angular power spectra of
the various components of unpolarised Galactic foreground emission
appears to be well described by power law spectra with a spectral
index between 2 and 3.  Below, we model the angular power spectra of
the different components of the Parkes data with power-laws of the
form:
$$
C_l^{X} = k l^{-\alpha_X}
$$
where $l$ is the multipole order and $X$ may stand for $T$, $L$, $E$,
$B$, $e$ and $b$, with $e$ and $b$ indicating the E-B transform of the
sinus and cosinus of the polarisation angle. 

The limited sky coverage of the survey and the FWHM resolution of 10.4
arcmin limit our spectral analysis to the multipole range
$l=40$--250. Sect.~\ref{sec:sim} describes Monte Carlo simulations
that were made to assess the effects of reduced sky coverage and
finite resolution.

\subsubsection{Total and polarised intensity}

\begin{figure}[]
 \begin{center} \leavevmode \epsfig{file=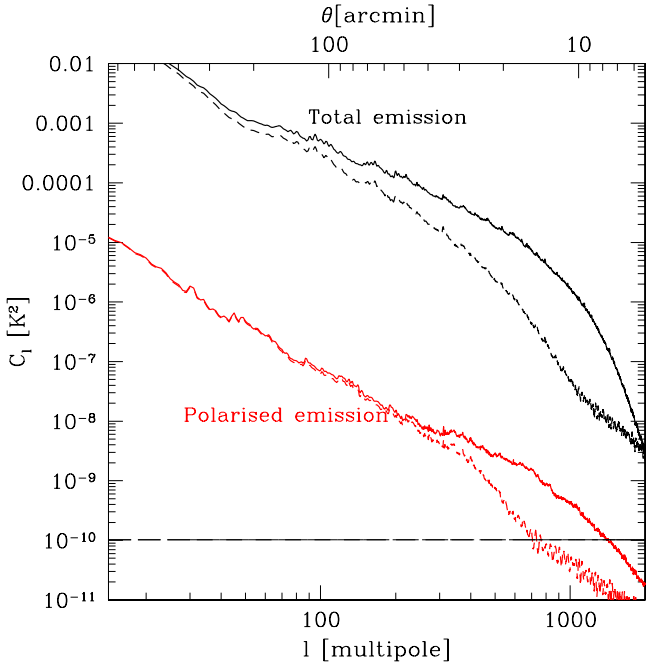,
   width=9.0cm, clip=} 

	\caption{The global angular power spectra of $T$ and
   $L=\sqrt{Q^2 + U^2}$ for the Parkes survey (continuous lines). The
   power spectra of $T$ and $L$ after median filtering are also shown
   (dashed lines). Point source suppression by the filtering process
   is responsible for the steepening of the intensity spectrum. Such
   steepening is not observed in the spectrum of polarised emission
   (up to $l\sim150$). The horizontal line gives the nominal noise
   level.  The multipole order $l$ corresponds to a typical angular scale
   of $\theta = \pi/l$.}

	\label{fig:PSip_ipmf} 
    \end{center}
\end{figure}

The angular power spectrum of the $T$ channel of the Parkes survey is
shown in Fig.~\ref{fig:PSip_ipmf}.  In the multipole range $l
=$40--250 a linear least-squares fit to this spectrum gives:
$\alpha_{T} = 1.67 \pm 0.15$\footnote{Errors quoted in this paper are
estimated via Monte Carlo simulations unless otherwise stated}. In the
figure the angular power spectrum of the intensity of the polarised
signal $L$ is also shown. Linear least-squares fit to this spectrum
gives $\alpha_{L} = 2.37 \pm 0.21$.

From the images in Fig.~\ref{fig:parkes} point sources are clearly
identifiable  in the map of total emission while they are not 
identifiable in the polarised channels.
In order to verify whether point sources or noise at pixel scale is
affecting the spectra derived from the Parkes survey, we median
filtered the intensity map ($T$) and the derived polarised intensity
map ($L$) by convolving the data with a kernel of 9$\times$9
pixels. We have used this technique to derive the angular power
spectrum of diffuse emission from the large sky radio survey of
\cite*{Jonas98}. We showed that median filtering can be used to
effectively remove point sources and derive the angular power spectrum
of the diffuse component (\cite{Giardino01}).

The angular power spectrum of a population of point sources randomly
distributed in the sky is Poissonian, that is with constant power at
all $l$. Therefore, if point sources contribute significantly to the
temperature fluctuation in the field of sky, the power spectrum of
that field is expected to steepen when the point sources are removed
and only the diffuse components remain. The same happens if the signal
is dominated by instrumental random noise.  On the other hand if the
signal in the field is dominated by the diffuse component, median
filtering will not have any effects on the spectrum of the emission up
to the multipole order where the high-frequency fluctuations of the
emission are averaged out.  We estimated by Monte Carlo simulation
that, for a beam with a FWHM of 10.4 arcmin and a map with pixel size
of 3.4 arcmin, median filtering with a 9$\times$9 pixel kernel allows
the input angular power spectrum of a diffuse component (with a
spectral index of $\alpha = 2$) to be derived up to $l=150$ with an
error lower than 3\%.

The results of the analysis of the median filtered $T$ and $L$ maps
are also shown in Fig.~\ref{fig:PSip_ipmf}.

From the figure it is apparent that the spectrum of the total emission
steepens significantly upon the application of the filtering
process. The fitted spectral index values, in the $l$ range 40--150,
before and after median filtering are respectively $\alpha_{I} =
1.68\pm 0.03$ and $\alpha_{I}^{\rm mf}=2.05 \pm 0.03$ (only the formal
fit errors are given in this case\footnote{Since in this case we are
interested in the spectral indices' relative values before and after the
filtering, the formal fit errors are representative of the uncertainty
in this comparison}).  This indicates that a significant fraction of
the power spectrum at intermediate and high $l$ is due to the presence
of discrete signals: either a population of point sources or
instrumental noise at the pixel scale. In fact instrumental noise at
pixel scale is unlikely as the nominal noise level is well below the
spectrum derived for the total intensity.  We therefore interpret the
steepening observed as due to the removal of the point source
contribution.

Fig.~\ref{fig:PSip_ipmf} shows no comparable change in the slope (for
$l<150$) of the angular power spectrum of the derived $L$ map after
median filtering. The fitted spectral index values before and after
median filter are respectively $\alpha_L = 2.34\pm 0.03$ and
$\alpha_{L}^{\rm mf}=2.45 \pm 0.03$. 

This analysis implies that, up to multipole order $l = 150$, the
fluctuations in the polarised emission are of diffuse nature and they
are not due to the presence of discrete signals such us point sources
or instrumental noise at pixel scale.

Even though we can exclude contamination of the polarisation spectrum
by point sources and instrumental noise at pixel scale only up to
multipole $l=$150, in the following sections we derive the spectral
indices for the other components of the Parkes data over the entire
multipole range $l=40$--250 that can be directly probed given the
survey's sky coverage and angular resolution. The reader should note
from Fig.~\ref{fig:PSip_ipmf} that the slope of the power spectrum of
the unfiltered polarised emission does not obviously change between an
$l$ of 150 and 250, and therefore the fits given up to 250 should also
be consistent with those made only up to 150. Indeed, we have
explicitly established this.

\subsubsection{The $E$ and $B$ components}

\begin{figure}[!thbp]
 \begin{center} \leavevmode \epsfig{file=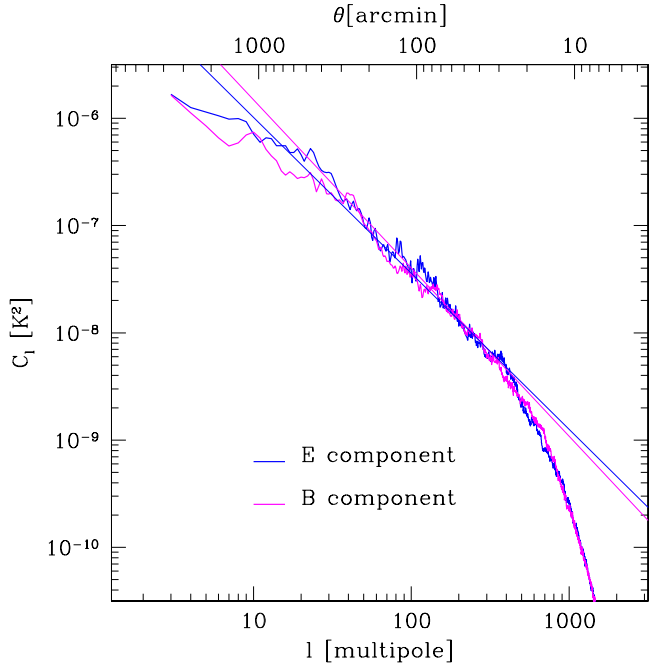, width=9.0cm,
   clip=} 
	
	\caption{The power spectra of the $E$ and $B$ components for
   the Parkes data. The linear least squares fits to the spectra in
   the multipole range 40--250 are also shown.}
   
	\label{fig:rawPSeb} 
  \end{center}
\end{figure}

The power spectra of the $E$ and $B$ components derived from the
Parkes survey are shown in Fig.\ref{fig:rawPSeb}. In the multipole
range $l =40$--250 linear least-squares fits to the spectra give
$\alpha_{E} = 1.57 \pm 0.12$ and $\alpha_{B} =1.45\pm 0.12$
respectively for $E$ and $B$.  As can be seen from the figure the
power spectra of the $E$ and $B$ components are very similar in slope
and amplitude. This is expected if the observed emission field does
not show clear patterns or symmetries (\cite{Zaldarriaga01}).

\subsubsection{The $E \times T$ component}

\begin{figure}[!thbp]
 \begin{center} \leavevmode \epsfig{file=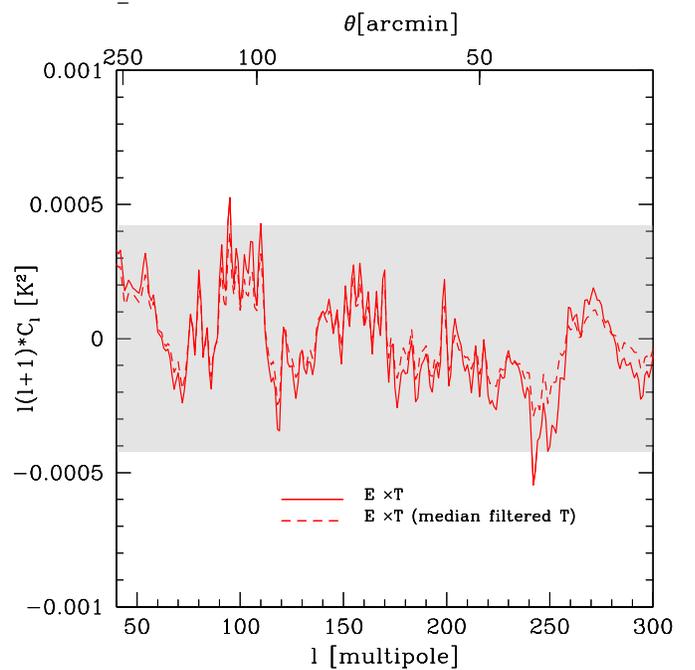,
   width=9.0cm, clip=} \caption{The cross-correlation $E \times T$ for
   the Parkes data. The grey area gives $\pm 3\sigma$
   from zero of the recovered $E \times T$ spectra
   for 10 simulations with input null $E \times T$ (see text).}
   \label{fig:rawPSet} \end{center}

\end{figure}

The cross-spectrum of the $E$ and $T$ components is
shown in Fig.~\ref{fig:rawPSet}.

In order to assess whether the computed correlation signal corresponds
to null correlation between $E$ and $T$ we performed a series of
simulations. We generated ten random Gaussian polarised fields using
the spectral indices and normalisation derived from the Parkes data to
define the power spectra of $T$, $E$ and $B$ and setting $C_{l}^{E
\times T}$ to zero. The same sky coverage and beam FWHM of the Parkes
survey were reproduced (see Sect.~\ref{sec:sim}).  

The $E \times T$ spectra derived from the ten simulations showed a
standard deviation around 0 of $\sigma_{l}=1.4\cdot
10^{-4}/l(l+1)~{\rm K}^2$.  The grey shaded region in
Fig.~\ref{fig:rawPSet} shows the area within $\pm 3\sigma_{l}$.  The
comparison of the data points with this area is consistent with the
absence of correlation between $E$ and $T$.  If a level of correlation
between $E$ and $T$ exists, this is below the level that can be probed
with the Parkes data, given the survey's sky coverage and beam FWHM.

 The power spectrum derived from $E$ and the
median filtered $T$ is also shown in the figure. The level of
correlation is hardly affected by the filtering process (up to the
multipole order $l = 150$ where the two spectra are directly
comparable), implying that the lack of correlation between the
$T$ and the $E$ component is not due to the presence of unpolarised
point sources (which only affect the $T$ map).

\subsubsection{The polarisation angle}

\begin{figure}[]
 \begin{center} \leavevmode \epsfig{file=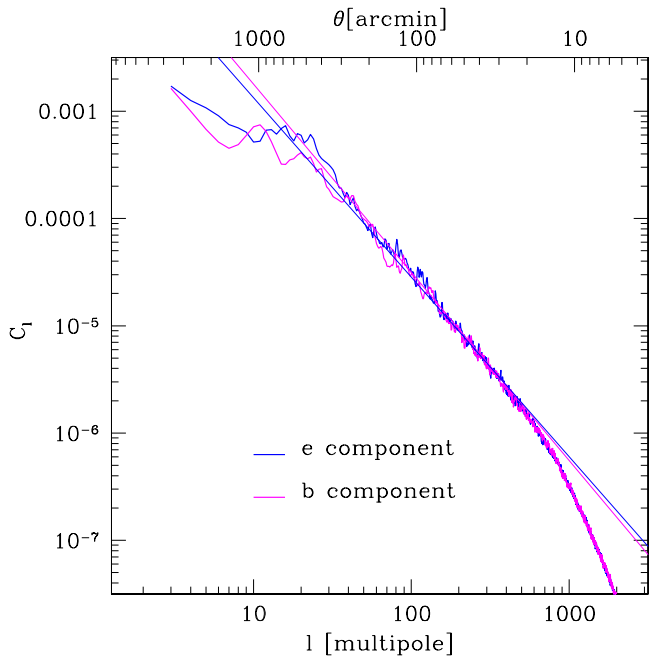,
   width=9.0cm, clip=} 
	
	\caption{The power spectrum of the E-B transform of
   $\cos(2\theta)$ and $\sin(2\theta)$, where $\theta$ is the
   polarisation angle. The transformed map are here indicated with $e$
   and $b$. The linear least squares fits to the spectra in the
   multipole range $l=40$--250 are also shown.}

   \label{fig:sincos} \end{center}
\end{figure}

In order to quantify the statistical properties of the polarisation
angle variations we constructed the maps of $\cos(2\theta)$ and
$\sin(2\theta)$, where $\theta$ is the polarisation angle.  These
maps are simply given by $Q/L$ and $U/L$ respectively. The sinus and
cosinus maps were then E-B transformed to obtain the rotationally
invariant quantities, $e$ and $b$. The power spectra of $e$ and $b$
were also derived by spin-weighted harmonic analysis. The result of the
analysis is shown in Fig.~\ref{fig:sincos}.

In the multipole range $l=40$--250, the power spectra of $e$ and $b$
are well described by power law spectra with fitted spectral
indices of $\alpha_{e} = 1.74 \pm 0.14$ and $\alpha_{b} = 1.69 \pm
0.13$ respectively.  These values are very close to the spectral
indices of the power spectra of the $E$ and $B$ components.

\section{Error assessment}
\label{sec:sim}
 
In order to verify the reliability of the power spectra derived from
the Parkes survey we performed a series of Monte Carlo simulations. We
generated 10 full-sky realisations of a random Gaussian polarised
field defined by four power spectra with shape and normalisation
similar to the ones of the Parkes data. For $T$, $E$ and $B$ we used
power-law spectra with spectral indices of $\alpha_T = 1.7$, $\alpha_E
= 1.5$, $\alpha_B = 1.5$ respectively, for $E \times T$ the spectrum
was set to zero. The 10 full-sky random realisations of these spectra
were convolved with a Gaussian beam with a FWHM of 10.4 arcmin and
sampled using a HEALPix tessellation with {\em nside}$=1024$.  The power
spectra for each realisation were then derived by performing 
spin-weighted harmonic analyses for both full-sky coverage
and the limited Parkes sky coverage.

\begin{figure*}[]
 \begin{center} \leavevmode \epsfig{file=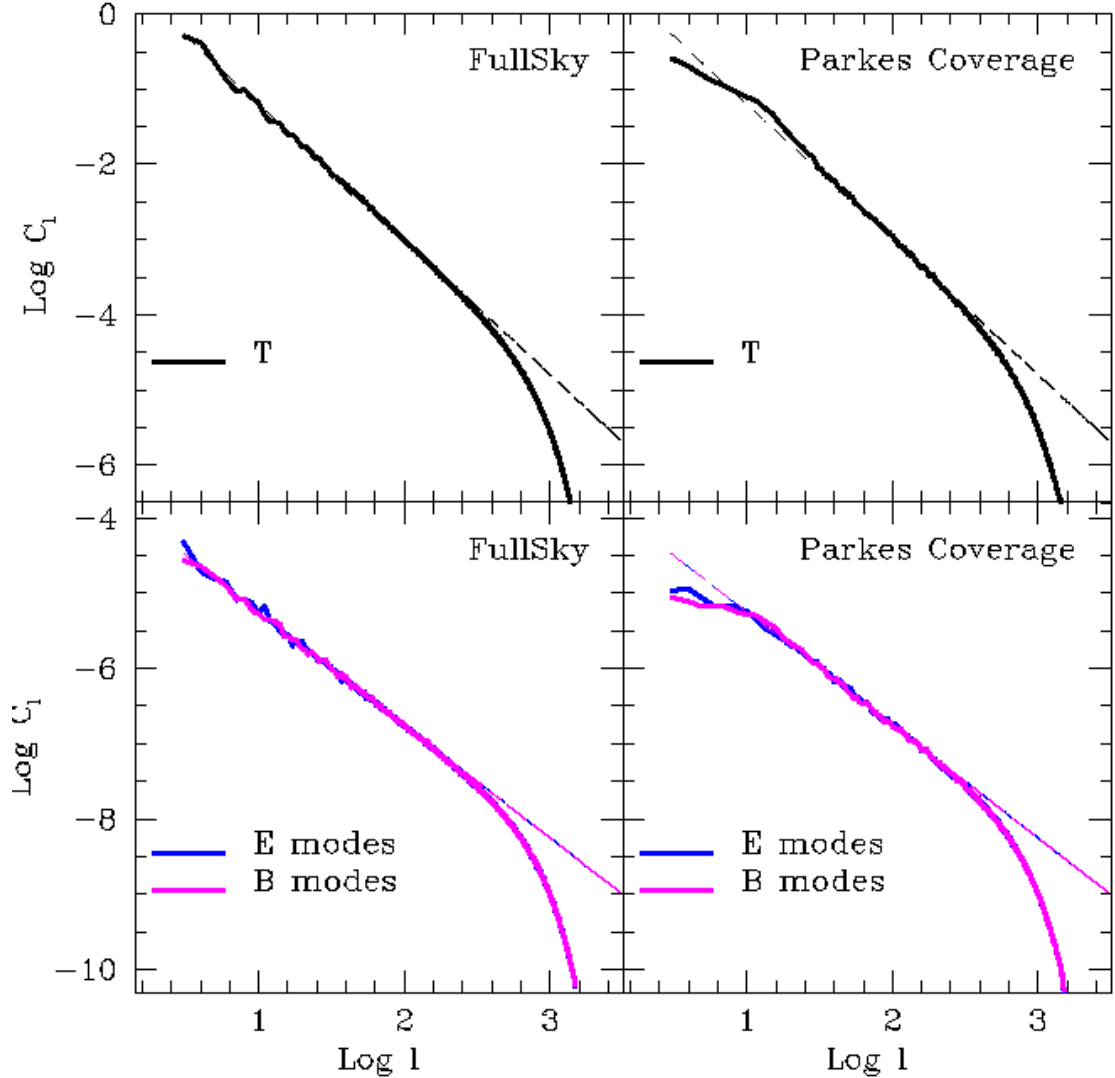,
 width=15.0cm, clip=} 

	\caption{The average spectra of $T$ (top) and of $E$ and $B$
   (bottom) obtained from the spin-weighted harmonic analysis of 10
   different realisations of a polarised Gaussian radiation field
   (continuous lines). The dashed lines indicates the input power
   spectra for $T$, $E$ and $B$. These are power law with spectral
   indices $\alpha_T = 1.7$, $\alpha_E = 1.5$, $\alpha_B = 1.5$ (very
   similar to the Parkes data). The input spectrum for $E\times T$ was
   set to zero.}

\label{fig:simulPS} 
\end{center}
\end{figure*}

The averaged power spectra from the 10 simulations are shown in
Fig.~\ref{fig:simulPS}. The top panels shows the derived average power
spectra of $T$ and the bottom panels the derived average power spectra of
$E$ and $B$. The two different cases of sky coverage are shown.  The
effect of the incomplete sky coverage is visible in the
spectra in the right panels where the average of the derived spectra
deviates significantly from the input spectra (the true value) for
$l < 40$. Above multipole order $l > 250$ the derived power spectra are
affected by the beam cut-off.

In order to quantify the accuracy of the recovery we have computed the
best-fit power law of the spectra obtained from the 10 realisations,
in the two different cases of sky coverage. The spectral indices of
these best-fit power laws can be compared with the true spectral
indices of the power spectra used as inputs for the 10
realisations. The standard deviation of the derived spectral indices
with respect to the true (input) values provides an estimate of the
standard error on the spectral index. The results are summarized in
Table~\ref{tab:recovCl}.  In the case of full sky coverage and
considering the multipole range $l = 3$--250, one can recover the input
spectral indices of $T$, $E$ and $B$ with an error of approximately
2\% (as can be seen by taking the ratio of the SD over the mean in
three columns of the table).  The limited sky coverage of the Parkes
survey implies that one can only fit a power law to the recovered
spectra in the multipole range $l = 40$--250. By considering the
spectrum in this limited multipole range the error on the recovered
spectral indices increases significantly even when using spectra
derived from full-sky maps (around 5\%, as taken from
Table~\ref{tab:recovCl}). The error is around 10\% when
recovering the indices of the $T$, $E$ and $B$ spectra from maps with
the limited sky coverage of the Parkes survey.

\begin{table*}
\begin{center}
\footnotesize
\begin{tabular}{|c|c c c c c c c|}
\hline
 ~ & ~ & $\alpha_T$ & ~ & $\alpha_E$ & ~ & $\alpha_B$ & ~ \\
\hline
$l$ range & Sky coverage  &  mean & SD  &  mean & SD  &  mean & SD \\
\hline
3$-$250 & Full  & 1.70 & 0.02 & 1.51 & 0.03 & 1.51 & 0.03\\
40$-$250 & Full & 1.76 & 0.08 & 1.56 & 0.08 & 1.55 & 0.08 \\
40$-$250 & Parkes & 1.78 & 0.15 & 1.57 & 0.12  & 1.55 & 0.12\\
\hline
\end{tabular}
\end{center}
\caption{Mean and standard deviation of the recovered spectral indices
from 10 different sky realisations of a random Gaussian field with
input spectra with spectral indices $\alpha_T=1.7$, $\alpha_E=1.5$,
$\alpha_B=1.5$ for $T$, $E$ and $B$. The spectrum of $E\times T$ was
set to zero. Two different
cases of sky coverage have been considered: full sky coverage and the
Parkes survey sky coverage.}
\label{tab:recovCl}
\end{table*}

\section{Comparison with previous results and discussion}
\label{sec:disco}

The spectral index of $\alpha_T= 1.67 \pm 0.15$ that we derived from
the Parkes data for the total intensity ($T$) is significantly lower
than the spectral index values derived from large sky radio survey
(\cite{Bouchet99}; \cite{Giardino01}). However this value refers to
the raw data which beside the diffuse emission contains a significant
population of point sources. After point source removal the index
steepens to $\alpha_T = 2.05 \pm 0.15$.  This value can be compared
with the value of $\alpha_T =1.9$ for the spectral index of the same
region of the sky of the Rhodes/HartRAO survey after point source
removal.

The spectral index of the Rhodes/HartRAO survey steepens with Galactic
latitude from $\alpha = 2.4$ when all the data are considered to
$\alpha = 2.9$ when only the data at galactic latitude $|b| > 20$\deg~
are taken into account (\cite{Giardino01}). Therefore it is not
surprising that when a region within Galactic latitude $|b| =8$\deg~
is considered the derived spectral index is flatter.  The reason for
this is probably  a combination of the contributions from
supernovae remnants and from the high concentration of diffuse
H\,{\sc ii} emission regions in the Galactic plane.

As earlier noted, the Parkes survey was previously
analysed by \cite*{Baccigalupi01} and \cite*{Tucci00}.  
Baccigalupi et al.  restricted their analysis to the total intensity
and the polarised intensity data ($L$) while \cite*{Tucci00} present
results for the power spectra of $T$, $E$ and $B$, but not $L$.

\cite*{Baccigalupi01} obtained an average value of $\alpha_{L} = 1.7
\pm 0.2$ for the spectral index of the angular power spectrum of the
polarised intensity of the twelve patches in the $l$-range 100--800.
This value is significantly lower than the value of $\alpha_{L} =2.37
\pm 0.21$ that we derived for the multipole range $l=40$--250.
However, there are two issues to consider:
\begin{enumerate}
  \item the angular power spectrum of the unfiltered $L$
    (Fig.~\ref{fig:PSip_ipmf}) shows a flattening of the slope
    around $l =300$. 
  \item in order to analyse the Parkes data beyond $l \sim 250$ where the beam
    suppression becomes appreciable, assumptions about the noise amplitude
    and distribution are necessary.
\end{enumerate}

If we perform a linear least squares fit to $C_l^L$ in the multipole
range $l=$100--800, assuming purely Gaussian random noise and 
the nominal noise level given in \cite*{Duncan97},
we obtain $\alpha_{L, l=100-800} = 1.75$, in close agreement with the
value derived by Baccigalupi et al.. On the other hand, if we take into
account the presence of the slope break and fit the curve with two
power laws, we obtain $\alpha_{L, l=40-250} = 2.35$ and $\alpha_{L,
l=250-800} = 1.64$, indicating a flattening of the spectrum at $l
>250$.  

Clearly, Baccigalupi et al. derived a flatter spectral index
by fitting the spectrum of $L$ in the $l$-range 100--800
without accounting for the slope break. 
As to the reason why their analysis did not see evidence
for a change in slope of the power spectrum of $L$,
it is likely that by analysing small patches, the derived
individual spectra are noisier, thus rendering it more difficult 
to detect such a change.

Unfortunately, assessing the significance of this flattening is
difficult precisely because the analysis for $l > 250$  is sensitive
to assumptions about the noise amplitude and its distribution.

\cite*{Tucci00} derived the average angular power spectra of the $E$
and $B$ components for the twelve patches and concluded that in the
multipole range $36 \le l \la 1000$ they are both well approximated by
power laws with spectral index $\alpha \simeq 1.4$. This value is very
similar to the values of $\alpha_{E} = 1.57\pm 0.12$ and $\alpha_{B}
=1.45\pm 0.12$ that we derived for the $E$ and $B$ components
respectively.

The angular power spectrum of $e$ and $b$, the E-B transform of the
sinus and cosinus of the polarisation angle, were first derived in the
present work.  In the multipole range $l=40$--250 the power spectra of
the $e$ and $b$ components of the Parkes data are well described by
power laws with fitted spectral indices of $\alpha_{e} = 1.74 \pm
0.14$ and $\alpha_{b} = 1.69 \pm 0.13$. These values are very close to
the spectral indices derived for the $E$ and $B$ components.
Fluctuations in $e$ and $b$ are only determined by
variations in polarisation angle.  Therefore this analysis shows that
fluctuations in the $E$ and $B$ components are mostly determined by
variations in the polarisation angle rather than polarisation intensity,
which, over this multipole range, is characterised by a significantly
steeper power spectrum.

Finally, we did not find evidence of correlation between $E$ and $T$
in the Parkes data.  This lack of correlation can be interpreted
as a direct consequence of the fact that fluctuations in the $E$
components are mostly determined by changes in the polarisation angle,
which $T$ remains unaffected by.

Since we are interested in using these results to construct a toy
model of the full-sky synchrotron polarisation at high frequencies
(30-100 GHz), there are two fundamental questions that we need to ask.

{\it i}) What is the effect of Faraday rotation on the Parkes data?

As a result of propagation through an ionized medium, the direction of
linear polarisation is rotated by an angle proportional to the
radiation wavelength: $\phi = {\rm RM}\lambda^2$.

\cite*{Gaensler01} have made high resolution (1 arcmin)
polarimetric observation of a 28$~{\rm deg^2}$ region in the Parkes
field (325\deg.5$\le l \le$332\deg.5, $-0$\deg.5$\le b \le$3\deg.5) at
1.4 GHz.  For this region, they derive a mean RM over the entire
region of $-12.9\pm 0.1~{\rm rad~m^{-2}}$. This at 2.4 GHz corresponds
to approximately $11^{\circ}$. However, the RM values span from 10 to
1000~${\rm rad~m^{-2}}$, with typical uncertainties in RM measurements
of $\pm 20-40~{\rm rad~m^{-2}}$.  At a frequency of 2.4 GHz and
resolution of 10.4 arcmin, a linear gradient of 5 ${\rm rad~m^{-2}
{arcmin^{-1}}}$ in a foreground screen can significantly depolarise
background radiation (\cite{Sokoloff98}) and affect the measured
variation of polarisation angle. Beam depolarisation is also
significant at this frequency and resolution for a level of dispersion
in RM of 30 ${\rm rad~m^{-2}}$ (e.g. \cite{Tribble91}).

\cite*{Gaensler01} conclude that the brightest polarised features seen
in this direction of the Galactic plane most
likely represent intrinsic structure in the source of emission, but fainter
structures are best explained as being imposed by Faraday rotation on a
uniformly polarised background by foreground
material. 

This is likely to be the case also at 2.4 GHz and implies that a
significant fraction of the angular variations which determine the power
spectra of $E$ and $B$ for the Parkes data are likely to be induced by
Faraday rotation.  This scenario would predict that the angular power
spectra of $E$, and $B$ of synchrotron emission derived from the
Parkes survey may be flatter than if they were derived from
observations at frequencies higher than 10 GHz, where Faraday effects
are negligible (\cite{Beck00}).



{\it ii}) How representative are the results here derived on the
angular power spectrum of linearly polarised emission at low Galactic
latitudes of what may happen at high Galactic latitudes?

\cite*{Duncan97} show that over much of the longitude range covered by
the survey, there exists a background polarised component which
appears to be independent of latitude (over the range of latitudes
covered by the survey). Besides the Parkes data, \cite*{Baccigalupi01}
also analysed five 10\deg-patches from the Northern galactic plane
survey by \cite*{Duncan99} at 2.7 GHz and three patches of different
sizes at medium Galactic latitudes ( $2^{\circ}.5 \le b \le
17^{\circ}.5$) from the survey of \cite*{Uyaniker99} at 1.4 GHz. By
averaging the derived spectral indices for the polarised emission from
all twenty fields they obtain $\alpha_{L, l=100-800} =1.8\pm
0.3$. They concluded that synchrotron polarisation maintains
essentially the same statistical properties up to Galactic latitude
$b\sim 10$\deg.

However, by comparing high Galactic latitude fields with
observations of the Galactic plane at 1.4 GHz \cite*{Gray99} find that
the magneto-ionic medium which induces the Faraday rotation on
polarised emission appears to be concentrated in the disc component
of the Galaxy ($b \la 10$\deg). This would imply that higher
latititude observations of synchrotron polarised emission at low
frequencies and for similar angular resolution may yield steeper
angular power spectra.

Indeed, if there is a variation of the slope with Galactic latitude one
may expect this change to be in line with what is observed for the total
emission, for which the angular power spectrum steepens with Galactic
latitude. Moreover, as pointed out by \cite*{Davies99}, the magnetic
field pattern may be more ordered at high Galactic latitudes,
resulting in less small-scale structure.


\section{Full-sky synthetic maps of synchrotron polarisation}
\label{sec:toys}

As a starting point for creating a toy model of Galactic synchrotron
polarisation at high frequencies we assume that the polarised
component of synchrotron emission in a given direction is proportional
to the unpolarised intensity, which in this case is expressed in terms
of brightness temperature, $T$:
\begin{eqnarray}
Q = f T \cos(2 \theta) \\
U = f T \sin(2 \theta) \nonumber
\label{eq:toyformula}   
\end{eqnarray}
where $f$ is the fraction of polarised emission (or polarisation
degree) and $\theta$ is the polarisation angle.   As we have seen,
at the frequency of Parkes observations this is clearly not a valid
assumption, but it can be an acceptable starting point at frequencies
higher than 10 GHz, where Faraday effects (of rotation and
depolarisation) may be neglected.

The brightness temperature spectrum of synchrotron emission is well
described by a power law,
\begin{equation}
T(\nu) \propto \nu^{-\beta}
\label{eq:frequencydep}   
\end{equation}
where $\beta$ is referred to as the frequency spectral index.  In the
absence of non-uniform magnetic fields, the fraction of polarised
radiation $f$ is related to the spectral index (\cite{Cortiglioni95}):
\begin{equation}
f = \frac{3 \beta -3}{3 \beta -1}
\label{eq:poldegree}   
\end{equation}

Therefore in order to generate synthetic full-sky polarisation maps of
synchrotron emission at a given frequency, we need at least three
basic ingredients: a full-sky map of the total synchrotron emission at
a ``template'' frequency, a full-sky map of the {\em frequency}
spectral index and a model for the polarisation angle. The full-sky
map of the spectral index  will be used to generate $f$ in a given
direction and to extrapolate the total emission ``template'' to the
other frequencies where $Q$ and $U$ are generated according to Eq.~1.

\subsection{Synchrotron total intensity (T)}

The 408-MHz survey of \cite*{haslam} provides, up to date, the best
image of the total intensity of the full-sky Galactic synchrotron
emission. At this low frequency the synchrotron emission dominates the
Galactic diffuse signal (\cite{Beuermann85}; \cite{lawson}).  We used
the 408-MHz survey processed by D.P. Finkbeiner, M. Davis and
D. Schlegel (private communication). They have removed the point
sources and de-striped the map by applying a Fourier filtering
technique. Hereafter we will refer to this processed version of the
408-MHz survey as to the Cleaned 408 MHz map.

The angular resolution of the 408-MHz survey is 0.85\deg. The 
angular resolution of Planck will vary from the 33 arcmin FWHM in the
30 GHz channel to 5 arcmin FWHM in the 857 GHz channel. Since it is
conservative to assume that synchrotron emission will present
structure also at the smaller angular scale, we have superimposed
artificial fluctuations at angular scale $\la 1$\deg~ on the original
map.  

The map of artificial fluctuations was obtained from a full-sky
realisation of a Gaussian field having a power law spectrum with index
$\alpha =3$ for $l > 150$ and vanishing to zero for $l<150$. Before
being added onto the Cleaned 408 MHz\footnote{convolved with a beam
with FWHM of 0\deg .9} the fluctuation map is weighted by the Cleaned
408 MHz map itself\footnote{A Gaussian realisation of a power law
spectrum with index $\alpha=3$, for $l>150$ and vanishing to zero for
$l<150$ has (by definition) power spectra with the same amplitude for
all regions of the map. Such a map of artificial fluctuations (let us
call it $F$) cannot be added onto the Cleaned 408 MHz (let us called
this map $H$), where fluctuation amplitude for ($l<150$) is a strong
function of the position of the sky. By multiplying $F$ with $H/{\rm
max}(H)$ (``weighting'') one obtains a map of fluctuations (at $l>150$)
for which the amplitude varies as a function of their position in a
manner consistent with the Haslam map}. This operation can be viewed
as an extrapolation of the spectrum of the Cleaned 408 MHz map beyond
$l=150$ with a power law spectrum with index $\alpha =3$.  A power law
spectrum with spectral index $\alpha =3$ is similar to that derived
for the total synchrotron emission at high galactic latitude
($|b|>20$\deg~) from the analysis of the three large sky radio survey
up to $l \sim 100$ (\cite{Bouchet99}; \cite{Giardino01}).

\begin{figure*}[!thbp]
 \begin{center} \leavevmode \epsfig{file=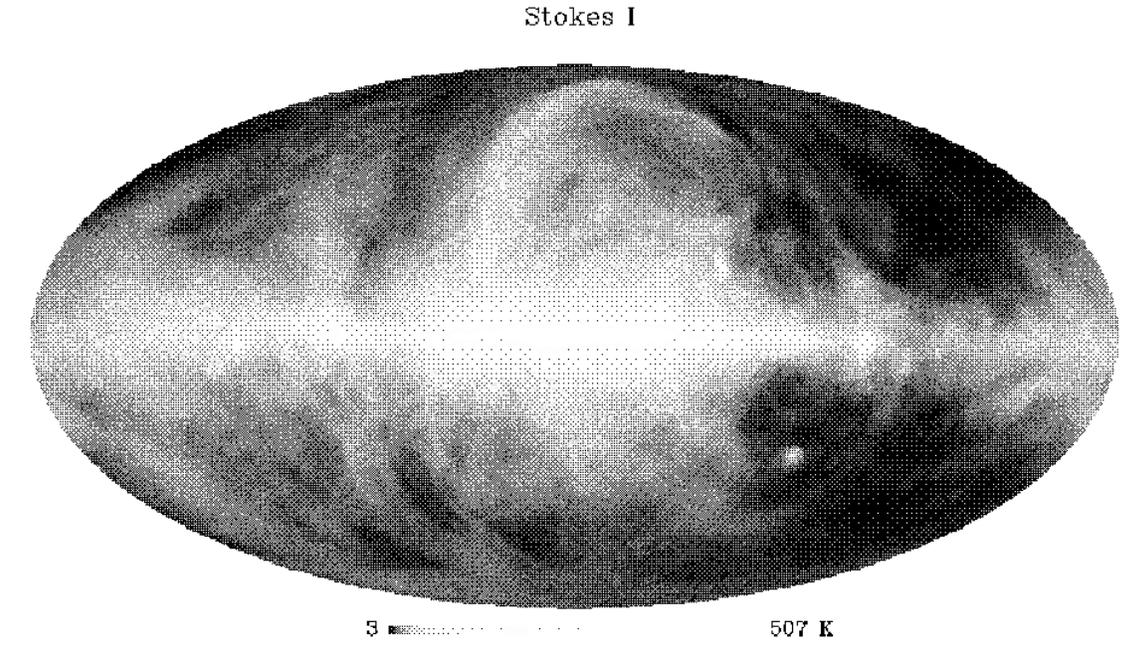, width=15cm,
   clip=} \caption{Mollweide projection of the Cleaned 408 MHz map
   with extrapolated fluctuations (i.e. the template for the Stokes
   parameter $T$ of the synthetic polarisation maps). The map is in
   Galactic coordinates. The color scale has been histogram
   equalised.}  \label{fig:Imap} \end{center}
\end{figure*}

The Cleaned 408 MHz map with extrapolated fluctuations provides us
with a realistic image of the full-sky total intensity of synchrotron
emission at 408 MHz up to the degree angular scale and an estimate of
how this emission may look like at the smaller unobserved angular
scales. We use this map as the template for the Stokes $T$ channel
of the synthetic polarisation maps. The template at 408 MHz is shown
in Fig.~\ref{fig:Imap}.

\subsection{Synchrotron frequency spectral index}

Besides the 408-MHz survey the other two high quality, large sky radio
surveys available to date are: the northern sky survey of
\cite*{reich} at 1420 MHz and the more recent southern sky survey of
\cite*{Jonas98}.  The sky coverage of the 1420-MHz survey and of the
2326-MHz survey are complementary to each other.  The two surveys
overlap over the Galactic declination range $ -20^{\circ}< \delta <
32^{\circ}$ in the right ascension interval $180^{\circ}\le RA \le
360^{\circ}$ and over Galactic declination range $-20^{\circ}< \delta
< 13^{\circ}$ in the right ascension interval $0^{\circ}< RA <
180^{\circ}$.  Therefore we can combine the 408-MHz survey and the
1420-MHz survey in order to derive a map of the spectral index of
radio emission in the northern sky and the 408 MHz and the 2326 MHz
survey in order to derive a map of the spectral index in the southern
hemisphere.  We obtained three homogeneous maps in the following way:
we took the surveys, as they are publicly available, resampled them
into a HEALPix tessellation with {\em nside}$=256$ and then median
filtered with a box kernel of 9$\times$9 pixels in order to suppress
the point source signal.

As we wish to determine the spectral index of the Galactic emission
only, the extragalactic emission must first be subtracted. This is
made up of two isotropic components, the CMB and the sum of unresolved
extragalactic background. The CMB brightness temperature is 
2.73~K. A phenomenological expression to estimate the integrated
brightness temperature of extra galactic unresolved radio sources at
different frequencies is given by \cite*{lawson}.  Lawson uses it to
estimate this contribution in the 408 MHz and in the 1420 MHz
surveys. We have adopted it to evaluate this contribution at 2326
MHz. The total amount of extragalactic contribution for the three
surveys together with basic information about their uncertainties is
summarised in Table~\ref{tab:surveysdata}. For the 1420 MHz
survey we adopted a baseline correction of $-$0.13 K (\cite{lawson}).

\begin{table*}
\begin{center}
\footnotesize
\begin{tabular}{|c| c c c c c|}
\hline
Survey  & Scale  &  Base level  & Extragalactic  & Baseline & Reference\\
Frequency (MHz) & error (per cent) & error (K) & background (K) & correction (K) & ~\\
\hline
408 & 5 & 3 & 5.92 & - & Lawson~et~al.~(1987)\\
1420 & 10 & 0.6 & 2.83 & $-$0.13 & Lawson~et~al.~(1987)\\
2326 & 5 & 0.080 & 2.75 & - & Jonas~et~al.~(1998)\\
\hline
\end{tabular}
\end{center}
\caption{Characteristics of the three radio continuum surveys that we
used to produce a full-sky map of the radio brightness temperature
spectral index.}
\label{tab:surveysdata}
\end{table*}

After subtraction of the extragalactic contribution, the brightness
temperature frequency spectral index $\beta_{\nu1 / \nu2}$ of the
Galactic radiation can be calculated. It is simply given by:
\begin{equation}
\beta_{{\nu_1}/{\nu_2}}= -\frac{T_{\nu_1}/T_{\nu_2}}{\log(\nu_1/\nu_2)}
\label{eq:beta}   
\end{equation}

Over the whole observed region the average value of the spectral index
obtained by combining the 408 MHz and the 1420-MHz survey is $\langle
\beta_{408/1420} \rangle = 2.78 \pm 0.17$. By combing the 408 MHz and
the 2326-MHz survey one obtains an average value of the spectral index
of $\langle {\beta}_{408/2326} \rangle=2.75 \pm 0.12$, over the whole
region observed. Here, the error is only the formal rms of the
spectral index maps and does not account for the systematic
uncertainties.  The mean values for $\beta$ agree well with the value
found by \cite*{Platania98} of $\langle {\beta}_{408/3800}\rangle =
2.73\pm 0.08$.

All the surveys are affected by scan-to-scan baseline errors. When the
ratio of brightness temperatures at two different frequencies is
calculated the effect of these residual errors is enhanced.  Because
of this, the map of $\beta_{408/1420}$ and the map of
$\beta_{408/2326}$ do not match in the overlapping region. We tried
various techniques to look for a baseline offset for one of the three
input maps that would minimise the differences between the two
spectral index maps. However discontinuities between the two spectral
index maps always remained apparent. 

In order to construct the full-sky map of the spectral index,
therefore, we used $\beta_{408/1420}$ for regions of the sky at
declination $\delta > 13$\deg~ and $\beta_{408/2326}$ for regions at
declination $\delta < -20$\deg.  In the overlapping region of the 1420
and 2326-MHz surveys ($ -20^{\circ} \le \delta \le 13^{\circ}$) we set
the spectral index to:
$$
\beta_{\rm overlap} = g \beta_{408/1420} + h \beta_{408/2326}
$$
where $g = \exp[-0.5 (\delta - 13^{\circ})^2/(10^{\circ})^2]$ and
$h=1-g$.  This choice of the weights $g$ and $h$ ensures that the two
$\beta$-maps merge smoothly into each other.  The region of sky at
declination $\delta < 83$\deg~ remains un-observed at 2326
MHz. So we ``padded'' this region with the spectral index value
$\langle {\beta}_{408/2326} \rangle$.

\begin{figure*}[!thbp]
 \begin{center}
   \leavevmode \epsfig{file=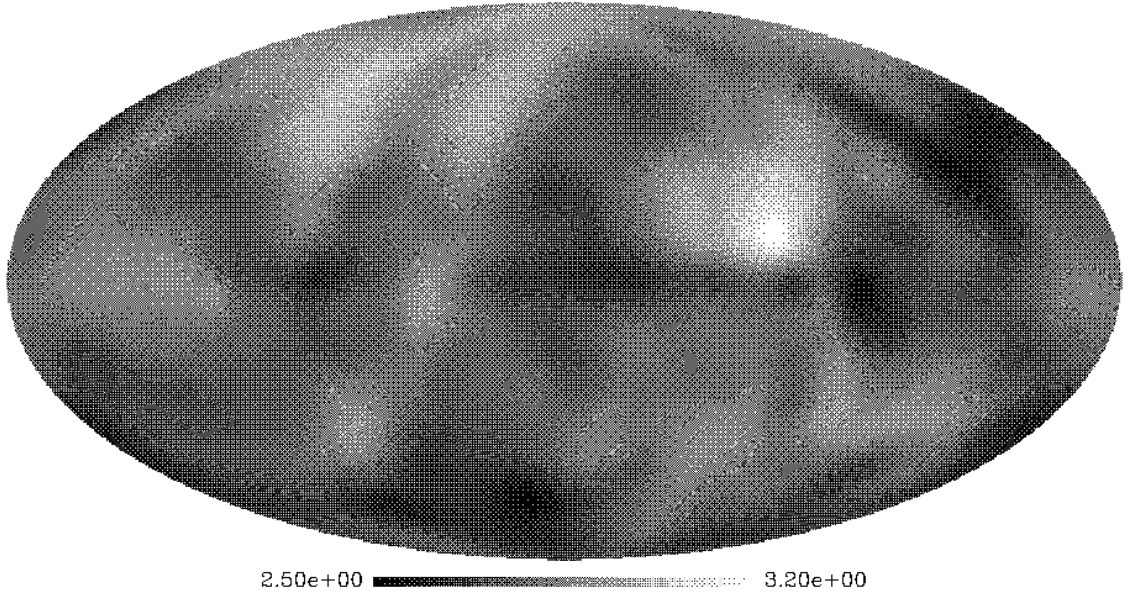, width=15cm, angle=180, clip=}
     \caption{Mollweide projection of the full-sky map of radio
   spectral index obtained by combining the 408-MHz survey, the
   1420-MHz survey and the 2326-MHz survey (see text). The map is in Galactic
   coordinates.}
     \label{fig:Betamap} 
 \end{center}
\end{figure*}

The spectral index map obtained in this way is however still affected
by the striation due to the scan-to-scan baseline errors in the input
data.  It is possible to remove the effect by Fourier filtering the
data but, because the scanning direction differs from one survey to
the other, one would not then be able to combine like with
like. Following \cite*{lawson} we adopted the simpler procedure of
convolving the data to a lower resolution, so that the baseline errors
are averaged sufficiently for the striation on the color-scale to
disappear. The final map of the spectral index is at a resolution of
10\deg~ (at FWHM), as shown in Fig.~\ref{fig:Betamap}.

\subsection{Synthetic polarisation angle}

We found that the power spectra of the E-B transform of the sinus and
cosinus of the polarisation angle of the Parkes data are described by
power laws with spectral indices of $\alpha_{e} = 1.74\pm 0.14$ and
$\alpha_{b} =1.69\pm 0.12$ in the multipole range $l=40$--250. One can
construct a full-sky map of a random angle having E-B transform of
its sinus and cosinus with the same spectra of the $e$ and $b$
component of the Parkes data in the following way.  Generate two
full-sky realisations of a Gaussian field characterised by a power law
spectrum with spectral index $\alpha =1.7$. Let us call these maps $x$
and $y$. Normalise $x$ and $y$ to the interval $[-1, 1]$. Generate the
full-sky map of the random angle by computing $\theta =
\frac{1}{2}\arctan(x/n,y/n)$, where $n= \sqrt{x^2+y^2}$.  We used the
random angle map obtained in this way as the 
polarisation angle of our toy model of Galactic synchrotron emission.

\begin{figure*}[!thbp]
 \begin{center}
   \leavevmode \epsfig{file=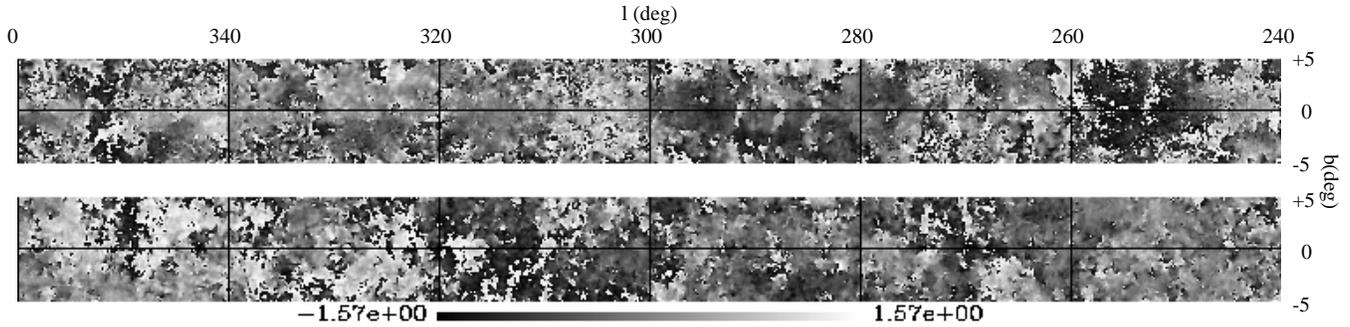, height=18cm, angle =-90}
     \caption{Polarisation angle from the Parkes survey (top) and from
   the synthetic map of synchrotron polarisation (bottom). Maps are in
   Galactic coordinates.}
     \label{fig:Angmap} 
 \end{center}
\end{figure*}

Fig.~\ref{fig:Angmap} shows the map of the polarisation angle derived
from the Parkes data and the map of the synthetic polarisation
angle for the same region of sky. The structure of the angle
fluctuations in the two maps looks similar.

\subsection{The synthetic maps at 30 and 100 GHz}

In order to generate the synthetic maps of synchrotron emission
and linear polarisation at higher frequencies the total intensity map
at 408 MHz has to be extrapolated. One can use the spectral index
map directly, assuming a fixed power law to higher frequencies, or use
it in conjunction with the local electron spectrum, as in
\cite*{Bennett92}, in order to derive maps of $\beta$ between any two
given frequencies and to attempt to allow for the observed steepening of $\beta$
with increasing frequency (\cite{lawson}; \cite{Banday90}).
 
For the local electron spectrum, we used the polynomial fit to the
spectral shape of electron measurements given in \cite*{Bennett92},
for energies $0.1< E({\rm GeV}) < 100$, and likewise a power law
$E^{-3.312}dE$ for $E > 100$ GeV. By interpreting the spatial
variation in $\beta$ (408-1420 MHz and 408-2326 MHz) as due solely to
variations in a synchrotron effective Galactic magnetic field, $B_{\rm
eff}$, we derived a full-sky map\footnote{nearly full-sky: only the
region at declination $\delta < 83$\deg~ remains un-observed at 2326
MHz} of $B_{\rm eff}$, with $\langle {B}_{\rm eff} \rangle = 2.0 \pm
2.3$~$\mu$G and a median value of 1.4~$\mu$G. This is remarkably close
to the values derived by \cite*{Bennett92} who did not have the data
from the 2326-MHz survey for the southern hemisphere.

The same expression for the local electron spectrum and the map of
synchrotron effective Galactic magnetic field were then used to obtain
the maps of the spectral index between 408 MHz and 30 and 100 GHz,
necessary for extrapolating the total emission map.  The
$\beta$ maps mean values are $\langle {\beta}_{0.408/30}\rangle =2.91
\pm 0.09$ and $\langle {\beta}_{0.408/100}\rangle =2.96 \pm 0.08$. The
value of $\langle {\beta}_{0.408/30}\rangle$ is consistent with the
upper limit of $\beta > 3$ between 408 MHz and 31.5 GHz obtained by
\cite*{Kogut96b} and the value of $\beta \sim 2.8$ between 1420 MHz and
19 GHz derived by \cite*{deOliveira-Costa98}.

\begin{figure*}[!thbp]
 \begin{center} \leavevmode \epsfig{file=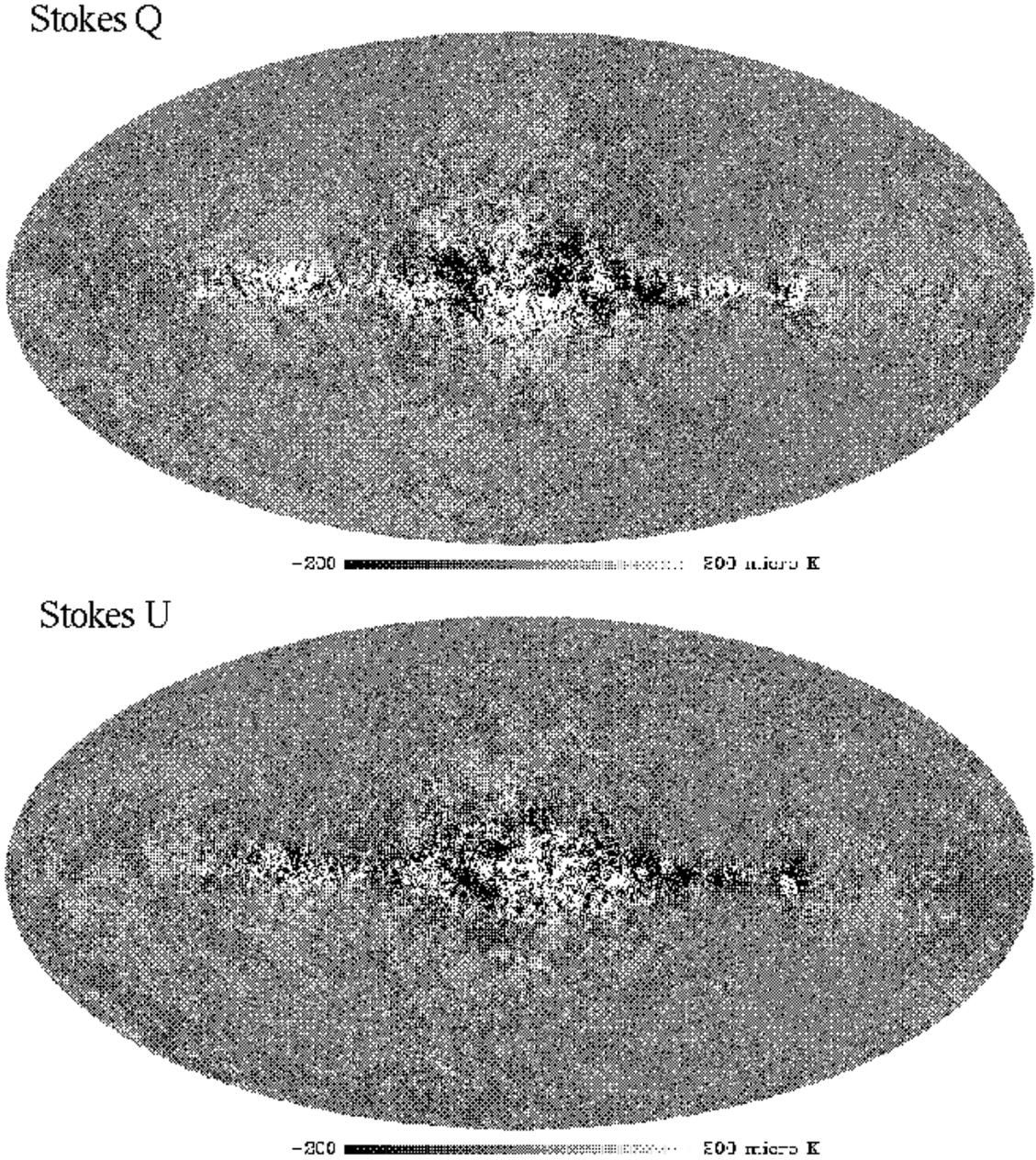, width=15cm, angle=0,clip=} 

	\caption{The $Q$ and $U$ channel of the synthetic map of
   Galactic synchrotron linear polarisation at 30 GHz. The map is in
   Galactic coordinates.}

	\label{fig:mapQU}
   \end{center}
\end{figure*}

The $Q$ and $U$ channels of the synthetic linear polarisation maps of
Galactic synchrotron emission at 30 and 100 GHz were then obtained by
combining the intensity map ($T$) extrapolated at 30 and 100 GHz with
the corresponding spectral index map and the random angle map,
according to Eq.~(1) and Eq.~(\ref{eq:poldegree}).  The $Q$ and $U$
channels at 30 GHz are shown in Fig.~\ref{fig:mapQU}.  The three
input maps are sampled into a HEALPix tessellation with a pixel size
of 6.9 arcmin ($nside=512$). Therefore also the synthetic maps of $T$,
$Q$ and $U$ are at a pixel resolution of 6.9 arcmin.

Because our toy model for $f$ is a very slowly varying function of the
position in the sky the maps of $Q$ and $U$ derived according to
Eq.~(1) have temperature fluctuations mostly determined by the
polarisation angle. Therefore, like for the Parkes data, the angular
power spectra of the $E$ and $B$ components of the synthetic linear
polarisation maps reflect the spectra of $e$ and $b$. They are
power-law spectra with indices $\alpha_E =  1.67\pm 0.09$ and
$\alpha_B=1.73 \pm 0.09$ over the $l$ range 40--250 ($\alpha_E =
1.59\pm 0.03$ and $\alpha_B=1.68 \pm 0.03$ for the $l$
range 3--250, that can be probed since in this case the power spectra
are derived from full sky maps). These values are very close to the
values derived for the spectra of the $E$ and $B$ components of the
Parkes data.

\subsubsection{Comparison of the power spectra with predictions for the CMB}

\begin{figure}[!thbp]
 \begin{center} \leavevmode \epsfig{file=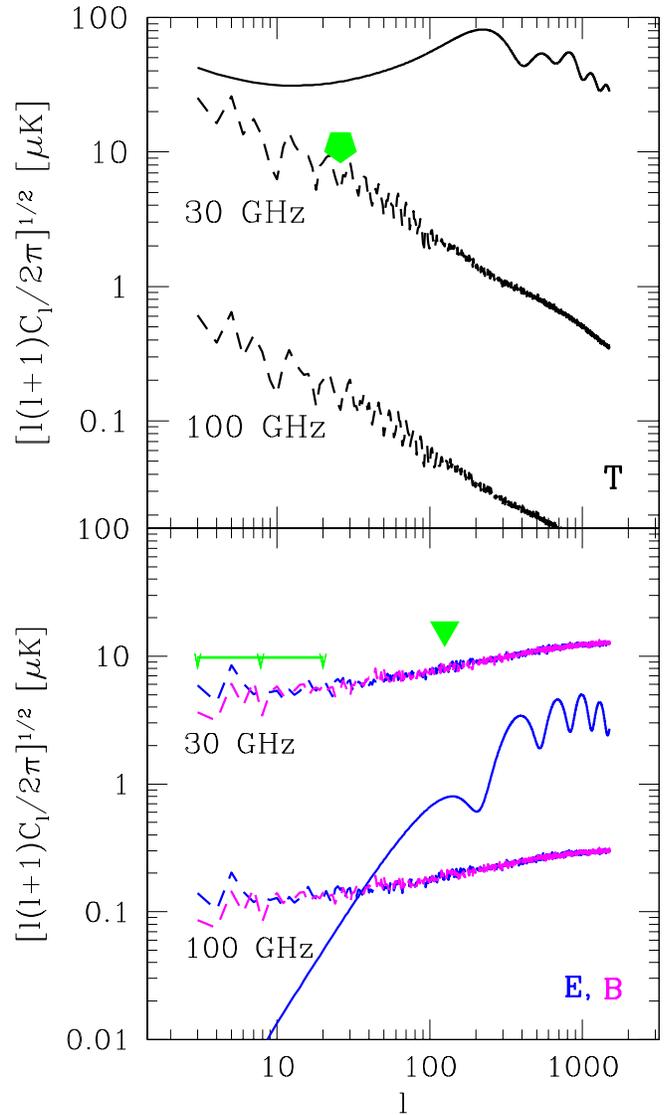, width=9.0cm,
   clip=} 
	
\caption{The power spectra of the $T$, $E$ and $B$ components of the
synthetic maps of Galactic synchrotron polarisation at 30 and 100 GHz
(dashed lines), for $|b|>20$\deg, compared with the signal expected
from the CMB (continuous line). The point with pentagonal shape
indicates the COBE DMR upper limit to synchrotron temperature
fluctuations at 31.5 GHz. The horizontal line is the upper limit to
polarised signal by Keating~et.al~(2001) in the frequency band
26$-$36~GHz. The triangular point is the upper limit to polarised
signal by Netterfield~et~al.~(1995) in the frequency band
26$-$46~GHz.}

	\label{fig:psIQU} 
 \end{center}
\end{figure}

For CMB studies one is particularly interested in the properties of
radio emission at high Galactic latitudes because this is where the
Galactic signal is weaker and where sensitive CMB measurements can be
made. Therefore, we computed the angular power spectra of the $T$, $E$
and $B$ components of the extrapolated maps with a Galactic plane
cut-off at latitude $|b|=20$\deg.  The results are shown in
Fig.~\ref{fig:psIQU}, where they can be compared with the signal
expected from the CMB. The CMB spectrum has been computed assuming a
flat inflationary model with purely scalar scale-invariant
fluctuations, vacuum energy $\Omega_{\Lambda} = 0.70$, cold dark
matter density $h^2\Omega_{\rm cdm}=0.12$ and baryon density
$h^2\Omega_{\rm b}=0.024$, (where $h = 0.7$ is the assumed Hubble
parameter), as derived by \cite*{Netterfield01} from combining the
latest CMB observations with constraints derived from measurements of
the large scale structure and results from recent measurements of type
Ia supernovae.

The top panel of Fig.~\ref{fig:psIQU} shows the angular power spectra
of the $T$ components of the two extrapolated maps and of the CMB.  Note
the absence of a perceivable change in the slope of the power spectra of the
$T$ components of the synthetic maps at $l=150$, the multipole order
at which artificial signal was added to the observed signal at 408
MHz\footnote{In order to prevent the aliasing noise that effects power
spectra with spectral index $\ga 3$ when a Galactic cut is applied we
used an apodysed Galactic cut off and the modified power spectra
definition given in Giardino et al. 2001, in which the terms $a_{l0}$
are set to zero}.

From the figure, it is apparent that at 100 GHz the contribution to
the total sky signal from high latitude synchrotron emission is
negligible. At 30 GHz synchrotron emission can be a significant
contribution to the sky anisotropy at scales larger than $\sim
20^{\circ}$, but it is more than an order of magnitude weaker than the
computed cosmological signal at degree angular scales where the first
``acoustic'' peak of the CMB is observed.  This confirms the results
discussed in \cite*{Giardino01} that were obtained with a different
approach using the index of the angular power spectra of the survey at
2326 MHz by \cite*{Jonas98} and the COBE DMR upper limit to
synchrotron temperature fluctuations at 31.5 GHz at the 7\deg~
scale. The COBE DMR upper limit (\cite{Kogut96b}) is indicated in the
figure by a pentagonal point.

The angular power spectra of the $E$ and $B$ components of the
synthetic maps are shown in the lower panel of Fig.~\ref{fig:psIQU}.
At small angular scale the power of temperature fluctuations in the
$E$ and $B$ components are mostly determined by change in the
polarisation angle rather than a change in the degree of
polarisation. This is the reason why temperature fluctuations in the
$E$ and $B$ components can become more intense than temperature
fluctuations in the $T$ channel.

The figure shows that at 30 GHz the high latitude synchrotron signal
may dominate the cosmological polarised signal over the entire
multipole range.  The extrapolated signal at 30 GHz is consistent with
the current upper limit on linearly polarised signal of 10 ~$\mu$K at
the 7\deg~ scale in the 26--36~GHz band (\cite{Keating01}) and with
the upper limit of 16~$\mu$K at degree scale in the 26--46~GHz band
(\cite{Netterfield95}).  These upper limits also refer to regions of
the sky at galactic latitudes $|b|>20^{\circ}$.

At 100 GHz the synchrotron signal can make a
significant contribution to the sky polarisation at large angular
scale but it is significantly weaker than the CMB $E$ component at
sub-degree angular scale. Therefore even using the most conservative
assumptions one can conclude that the contribution of synchrotron
polarisation to the sky signal will not hinder detection of the CMB
$E$ component, if this is present at sub-degree angular scale at the
intensity predicted by current cosmological models.

From the diagram the absence of the beam cut off in the spectra of the
synthetic maps is noticeable. These are artificial maps and therefore
they were generated at full resolution.

\subsection{Scope and limitations of the synthetic maps}
\label{sec:limitations}
 
The synthetic full-sky maps of the Stokes parameters $T$, $Q$ and $U$
of Galactic synchrotron emission at 30 and 100 GHz are a first
attempt to model the full-sky synchrotron polarisation at high
frequencies. The maps are needed as input for the simulation activity
that is carried out in preparation for the Planck missions.

To study the effects of polarised foreground on MAP observations of
the CMB, synthetic maps of polarised synchrotron emission have been
generated by \cite*{Kogut00}. They also assumed the polarised signal
to be proportional to the total intensity $T$, with the Stokes $Q$ and
$U$ components given by Eq.~2. For the polarisation degree, $f$, they
assumed a varying Gaussian field such that $\langle f \rangle = 0.1$.
They constructed the polarisation angle starting from Gaussian random
fields, $x$ and $y$, with flat power spectra\footnote{the fields were
then convolved with a Gaussian with varying FWHM, as the authors were
interested in investigating the effect on CMB observations of the
eventual presence of a coherence angle in polarised Galactic
emission}, which lead to $E$ and $B$ components whose spectra have
indices of $\alpha \simeq 0$.  The polarisation maps were only
generated for a fiducial frequency of 40~GHz.

We have constructed the synthetic maps by combining in a conservative
way many of the observational data currently available on synchrotron
total emission and polarisation.  The maps that can be obtained at
higher frequencies by extrapolating the total intensity
map with the spectral index map are consistent with available
observational upper limits on synchrotron emission and polarisation at
frequencies higher than 10 GHz. There remain however some
inconsistencies between the synthetic polarisation maps extrapolated
at 2.4 GHz and the Parkes data.

The synthetic polarisation maps reproduce the slope of the angular
power spectra of the $E$ and $B$ components of the Parkes survey, but
not their amplitude at 2.4 GHz. Moreover, the power spectra of $T$, $E
\times T$ and $L$  differ both in shape and normalisation to the ones
of the Parkes data.  This is because of the way that the toy model has
been constructed.  We derived information on the polarisation angle
(and the $E$ and $B$ power spectra) from the Parkes survey, while for
the intensity channel $T$ we used the data from the Haslam survey and
for the polarisation fraction $f$ we assumed the theoretical value
given by the synchrotron spectral index in a uniform
magnetic field. The reasons behind these choices are the following.

As already mentioned, the Haslam map provides the best image of the
total intensity of the full-sky synchrotron emission, to date. It is
natural to adopt it as a template for $T$ as opposed to using the $T$
channel of the Parkes survey. This is limited in sky coverage and
being at low Galactic latitude and a higher frequency contains a
higher fraction of thermal emission from ionised gas. 

The Parkes survey provides us with information about the statistical
properties of the polarisation angle, so we adopt it for this
purpose. Using the statistical properties of the polarisation angle
derived from a survey at decimetric frequency is a conservative
approach (i.e. it will unlikely lead to an underestimate of the level
of fluctuation of polarised synchrotron emission). This is because
Faraday rotation of a synchrotron background due to a foreground
``screen'' of ionised medium will tend to increase the level of
fluctuations in the observed polarisation direction.  On the other
hand Faraday differential rotation may decrease the polarised fraction
of the synchrotron signal at the Parkes frequency. Therefore using the
amplitude of the $E$ and $B$ components of the Parkes data may lead to
an underestimate of the polarisation fraction at higher frequencies
($\ga 10$ GHz).

Calculating the theoretical fraction of synchrotron polarisation by
assuming a completely uniform magnetic fields could be considered an
overly conservative approach because it is known that the interstellar
medium is pervaded by an isotropic random (``turbulent'') field.  In
the presence of a random magnetic field the observed level of
polarisation is given by $f_{\rm obs} = d\cdot f$. Here $f$ is the
intrinsic level of polarisation (given by Eq.~\ref{eq:poldegree}) and
$d$ is a coefficient ($\le 1$) which depends on the relative intensity
of the random field, $B_{\rm turb}$, and the field which is regular in
the whole resolution element, $B_{\rm reg}$(\cite{Burn66};
\cite{Beck98}).

Synchrotron polarisation observations at decimeter wavelength imply a
ratio of regular to total field strengths within about a kpc from the
Sun of $\langle B_{\rm reg}/B_{\rm turb}\rangle \simeq 0.6$ (\cite{Beck00}),
which, in the popular assumption of energy equipartition between cosmic rays
and magnetic fields would imply an average level of depolarisation of
$\langle d \rangle = 0.3$.  However the actual value varies
substantially between points and may depend on the size of the
resolution element. These values are derived from decimeter wavelength
surveys with degree-scale resolution. The level of depolarisation by
the random component of the magnetic field can be expected to decrease
as the sky is observed at higher resolution and the projected size of
the beam approaches the field coherence length.  Indeed convolution of
our synthetic maps with a Gaussian beam with a FWHM of 1\deg~ yields
an average level of depolarisation of $\langle d \rangle = 0.5$.


\section{Summary and Conclusions}
\label{sec:conclusion}

In order to obtain information about the statistical properties of
synchrotron polarised emission, we have derived the global angular
power spectra of the Parkes survey, a radio continuum and polarisation
survey of the Southern galactic plane at 2.4 GHz. The angular power
spectrum of the polarised intensity ($L$) is well approximated by a
power law with fitted spectral index $\alpha_L=2.37 \pm 0.21$ in the
multipole range $l=40$--250 that can be probed given the Parkes survey
sky coverage and angular resolution. We exclude that up to multipole
order $l = 150$, the spectral index value is affected by the presence
of discrete signals such us point sources or instrumental noise at
pixel scale.

The angular power spectra of the $E$ and $B$ components of the
polarised emission is significantly flatter than the spectrum of
$L$. They have fitted spectral indices respectively of $\alpha_{E} =
1.57\pm 0.12$ and $\alpha_{B} =1.45\pm 0.12$, in the multipole range
$l=40$--250. This is because temperature variations in the Stokes
channels $Q$ and $U$ of the data are mostly determined by variation in
polarisation angle rather than polarisation intensity.  The E-B
transform of the sinus and cosinus of the polarisation angle, the $e$
and $b$ components, are in fact also well described by power laws with
fitted spectral indices of $\alpha_{e} = 1.74 \pm 0.14$ and
$\alpha_{b} = 1.69 \pm 0.13$, in the same multipole range $l=40$--250.

We generated a full-sky map of a random polarisation angle having $e$
and $b$ components with power spectra with the same spectral indices
of the $e$ and $b$ components of the Parkes data. We combined this map
with a template of synchrotron emission at 408 MHz, derived from the
408-MHz survey, and with a full-sky spectral index map, that we
obtained by combining the 408-MHz survey with the 1420-MHz survey and
the 2326-MHz survey.  The simple recipe on how to combine the three
maps in order to derive a map of linear polarisation is given by
Eq.~(1) and Eq.~(\ref{eq:poldegree}).

In this way we have constructed synthetic maps of the Stokes
parameters ($T$, $Q$ and $U$) of Galactic synchrotron emission at
30 and 100 GHz. The angular power spectra of the $E$ and $B$
components of these synthetic maps have slope very similar to the $E$
and $B$ components of the Parkes data. Their polarisation intensity is
proportional to the total intensity $T$ as determined by the frequency
spectral index in the conservative hypothesis of a uniform magnetic
field.

We have  compared the angular power spactra of the synthetic maps
with predictions for the CMB and conclude that while direct
observations of the CMB $E$ component at 30 GHz may be precluded by
synchrotron polarisation, at 100 GHz, this component of Galactic
emission will not hinder the observation of cosmological polarisation,
if this is present at the predicted levels.

The synthetic maps of the linear polarisation of Galactic synchrotron
can be useful as a toy model to study the effect of the polarised
foreground on planned observations of the CMB.  Moreover they can
serve as the basis for the development of more advanced models which
incorporate data on the structure of the Galactic magnetic field. In
turn these refined models will be of fundamental importance in
interpreting the microwave polarisation data that will be provided by
the Planck and MAP satellite missions and the SPORT experiment.

\medskip

The synthetic maps of the Stokes parameters $T$, $Q$ and $U$ of
Galactic synchrotron emission at 30 and 100 GHz together with the
full-sky spectral index map are available for down loading at {\tt
ftp://astro.esa.int/pub/synchrotron}.

\begin{acknowledgements}
 We thank D.P. Finkbeiner, M. Davis and D. Schlegel for providing us
 with the Cleaned 408 MHz map, E. Hivon for helpful discussions,
 A.R. Duncan, R. F. Haynes, K.L. Jones and R.T Stewart for making the
 Parkes survey publicly available and U. Seljak \& M. Zaldarriaga for
 their CMBFAST software, which was used to generate the CMB angular
 power spectra.  The HEALPix analysis package
 (http://www.eso.org/science/healpix) was used extensively throughout
 this paper. We gratefully acknowledge very useful comments from an
 anonymous referee.

\end{acknowledgements}


\begin{thebibliography}{}

\bibitem[\protect\astroncite{{Baccigalupi} et~al.}{2001}]{Baccigalupi01}
{Baccigalupi} C., {Burigana} C., {Perrotta} F. et~al. 2001, A\&A in press

\bibitem[\protect\astroncite{{Banday} \& {Wolfendale}}{1990}]{Banday90}
{Banday} A.~J. \& {Wolfendale} A.~W. 1990, MNRAS 245, 182

\bibitem[\protect\astroncite{{Beck}}{1998}]{Beck98}
{Beck} R. 1998,
\newblock Galactic Foreground Polarisation, Proceeding of a Workshop held in
  Bonn,
\newblock eds. E.~M.~Berkhuijsen

\bibitem[\protect\astroncite{{Beck}}{2001}]{Beck00}
{Beck} R. 2001,
\newblock Galactic and extragalactic magnetic fields,
\newblock eds. R. Diehl et al., Space Science Reviews, Kluwer, Dordrecht

\bibitem[\protect\astroncite{{Bennett} et~al.}{1992}]{Bennett92}
{Bennett} C.~L., {Smoot} G.~F., {Hinshaw} et~al. 1992, ApJ Lett. 396, L7

\bibitem[\protect\astroncite{{Beuermann} et~al.}{1985}]{Beuermann85}
{Beuermann} K., {Kanbach} G. \& {Berkhuijsen} E.~M. 1985, A\&A 153, 17

\bibitem[\protect\astroncite{{Bond} \& {Efstathiou}}{1987}]{Bond87}
{Bond} J.~R. \& {Efstathiou} G. 1987, MNRAS 226, 655

\bibitem[\protect\astroncite{{Bouchet} \& {Gispert}}{1999}]{Bouchet99}
{Bouchet} F.~R. \& {Gispert} R. 1999, New Astronomy 4, 443

\bibitem[\protect\astroncite{{Burn}}{1966}]{Burn66}
{Burn} B.~J. 1966, MNRAS 133, 67

\bibitem[\protect\astroncite{{Cortiglioni} \&
  {Spoelstra}}{1995}]{Cortiglioni95}
{Cortiglioni} S. \& {Spoelstra} T. A.~T. 1995, A\&A 302, 1

\bibitem[\protect\astroncite{{Davies} \& {Wilkinson}}{1999}]{Davies99}
{Davies} R.~D. \& {Wilkinson} A. 1999,
\newblock in ASP Conf. Ser. 181: Microwave Foregrounds, ~77

\bibitem[\protect\astroncite{{de Oliveira-Costa}
  et~al.}{1998}]{deOliveira-Costa98}
{de Oliveira-Costa} A.~., {Tegmark} M., {Page} L.~A. \& {Boughn} S.~P. 1998, ApJ
  Lett. 509, L9

\bibitem[\protect\astroncite{{Duncan} et~al.}{1997}]{Duncan97}
{Duncan} A.~R., {Haynes} R.~F., {Jones} K.~L. \& {Stewart} R.~T. 1997, MNRAS 291,
  279

\bibitem[\protect\astroncite{{Duncan} et~al.}{1999}]{Duncan99}
{Duncan} A.~R., {Reich} P., {Reich} W. \& {F{\"u}rst} E. 1999, A\&A 350, 447

\bibitem[\protect\astroncite{{Fabbri} et~al.}{1999}]{Fabbri99}
{Fabbri} R., {Cortiglioni} S., {Cecchini} et~al. 1999,
\newblock in AIP Conf. Proc. 476: 3K cosmology,  194

\bibitem[\protect\astroncite{{Gaensler} et~al.}{2001}]{Gaensler01}
{Gaensler} B.~M., {Dickey} J.~M., {McClure-Griffiths} N.~M. et~al. 2001, ApJ
  549, 959

\bibitem[\protect\astroncite{{Giardino} et~al.}{2001}]{Giardino01}
{Giardino} G., {Banday} A.~J., {Fosalba} P. et~al. 2001, A\&A 371, 708

\bibitem[\protect\astroncite{{G\'orski} et~al.}{1999}]{Gorski98}
{G\'orski} K.~M., {Hivon} E. \& {Wandelt} B.~D. 1999,
\newblock in Proceedings of the MPA/ESO Cosmology Conference "Evolution of
  Large-Scale Structure" eds. A.J. Banday, R.S. Sheth and L. Da Costa, ~37

\bibitem[\protect\astroncite{{Gray} et~al.}{1999}]{Gray99}
{Gray} A.~D., {Landecker} T.~L., {Dewdney} P.~E. et~al. 1999, ApJ 514, 221

\bibitem[\protect\astroncite{Haslam et~al.}{1982}]{haslam}
Haslam C. G.~T., Salter C.~J., Stoffel H. et~al. 1982, A\&AS 47, 1

\bibitem[\protect\astroncite{{Hedman} et~al.}{2001}]{Hedman01}
{Hedman} M.~M., {Barkats} D., {Gundersen} J.~O., {Staggs} S.~T. \& {Winstein} B.
  2001, ApJ Lett. 548, L111

\bibitem[\protect\astroncite{{Jonas} et~al.}{1998}]{Jonas98}
{Jonas} J.~L., {Baart} E.~E. \& {Nicolson} G.~D. 1998, MNRAS 297, 977

\bibitem[\protect\astroncite{{Kamionkowski}}{1997}]{Kamionkowski97}
{Kamionkowski} M. 1997, Phys. Rev. D 55, 7368

\bibitem[\protect\astroncite{{Keating} et~al.}{2001}]{Keating01}
{Keating} B.~G., {O'Dell} C.~W., {de Oliveira-Costa} A. et~al. 2001, ApJ Lett.
  560, L1

\bibitem[\protect\astroncite{{Kogut} et~al.}{1996}]{Kogut96b}
{Kogut} A., {Banday} A.~J., {Bennett} C.~L. et~al. 1996, ApJ Lett. 464, L5

\bibitem[\protect\astroncite{{Kogut} \& {Hinshaw}}{2000}]{Kogut00}
{Kogut} A. \& {Hinshaw} G. 2000, ApJ 543, 530

\bibitem[\protect\astroncite{{Kosowsky}}{1996}]{Kosowsky96}
{Kosowsky} A. 1996, Ann. Phys 246, 49

\bibitem[\protect\astroncite{Lawson et~al.}{1987}]{lawson}
Lawson K.~D., Mayer C.~J., Osborne J.~L. et~al. 1987, {\it MNRAS} 225,
  307

\bibitem[\protect\astroncite{{Mandolesi} \& {Puget}}{1998}]{Mandolesi98}
{Mandolesi} N. \& {Puget} J.-L. 1998,
\newblock AAO for Planck HFI and LFI,
\newblock Paris: ESA

\bibitem[\protect\astroncite{{Netterfield} et~al.}{2001}]{Netterfield01}
{Netterfield} C.~B., {Ade} P. A.~R., {Bock} J.~J. et~al. 2001, ApJ submitted

\bibitem[\protect\astroncite{{Netterfield} et~al.}{1995}]{Netterfield95}
{Netterfield} C.~B., {Jarosik} N., {Page} L. et~al. 1995, ApJ Lett. 445, L69

\bibitem[\protect\astroncite{{Platania} et~al.}{1998}]{Platania98}
{Platania} P., {Bensadoun} M., {Bersanelli} M. et~al. 1998, ApJ 505, 473

\bibitem[\protect\astroncite{{Prunet} et~al.}{1998}]{Prunet98}
{Prunet} S., {Sethi} S.~K., {Bouchet} F.~R. et~al. 1998, A\&A 339, 187

\bibitem[\protect\astroncite{Reich \& Reich}{1986}]{reich}
Reich P. \& Reich W. 1986, A\&AS 63, 205

\bibitem[\protect\astroncite{Rybicki \& Lightman}{1979}]{rybicki}
Rybicki G.~B. \& Lightman A.~P. 1979,
\newblock Radiative Process in Astrophysics,
\newblock Cambridge University Press

\bibitem[\protect\astroncite{{Sokoloff} et~al.}{1998}]{Sokoloff98}
{Sokoloff} D.~D., {Bykov} A.~A., {Shukurov} A. et~al. 1998, MNRAS 299, 189

\bibitem[\protect\astroncite{{Tribble}}{1991}]{Tribble91}
{Tribble} P.~C. 1991, MNRAS 250, 726

\bibitem[\protect\astroncite{{Tucci} et~al.}{2000}]{Tucci00}
{Tucci} M., {Carretti} E., {Cecchini} S. et~al. 2000, New Astr. 5, 181

\bibitem[\protect\astroncite{{Uyan{\i}ker} et~al.}{1999}]{Uyaniker99}
{Uyan{\i}ker} B., {F{\"u}rst} E., {Reich} W., {Reich} P. \& {Wielebinski} R.
  1999, A\&AS 138, 31

\bibitem[\protect\astroncite{{Wright}}{1987}]{Wright87}
{Wright} E.~L. 1987, ApJ 320, 818

\bibitem[\protect\astroncite{{Zaldarriaga}}{2001}]{Zaldarriaga01}
{Zaldarriaga} M. 2001, Phys. Rev. D submitted

\bibitem[\protect\astroncite{{Zaldarriaga} \& {Seljak}}{1997}]{Zaldarriaga97}
{Zaldarriaga} M. \& {Seljak} U. 1997, Phys. Rev. D 55, 1830


\end{thebibliography}
\end{document}